\documentclass[twocolumn,pra,showpacs,floatfix]{revtex4}
\pdfoutput=1
\usepackage{amsfonts, amsmath, amssymb}
\usepackage{graphicx, color}
\usepackage{verbatim}

\newcommand{\abs}[1]{\lvert#1\rvert}
\newcommand{\C}{_{\mathbf{C}}}
\newcommand{\I}{_{\mathbf{I}}}
\newcommand{\QC}{_{\mathrm{QC}}}
\newcommand{\kB}{\ensuremath{k_B}}
\newcommand{\ms}{\ensuremath{\mathrm{ms}}}
\newcommand{\nK}{\ensuremath{\mathrm{nK}}}
\newcommand{\pbf}{\ensuremath{\mathbf{p}}}
\newcommand{\rbf}{\ensuremath{\mathbf{r}}}
\newcommand{\Tc}{\ensuremath{T_{c}}}
\newcommand{\reff}[1]{(\ref{#1})}
\begin{document}

\title{Condensation and quasicondensation in an elongated three-dimensional Bose gas}

\author{Michael~C.~Garrett}
\affiliation{The University of Queensland, School of Mathematics and Physics, Queensland 4072, Australia}
\author{Tod~M.~Wright}
\email{todw@physics.uq.edu.au}
\affiliation{The University of Queensland, School of Mathematics and Physics, Queensland 4072, Australia}
\author{Matthew~J.~Davis}
\affiliation{The University of Queensland, School of Mathematics and Physics, Queensland 4072, Australia}

\begin{abstract}
We study the equilibrium correlations of a Bose gas in an elongated three-dimensional harmonic trap using a grand-canonical classical-field method.  We focus in particular on the progressive transformation of the gas from the normal phase, through a phase-fluctuating quasicondensate regime to the so-called true-condensate regime, with decreasing temperature.  Choosing realistic experimental parameters, we quantify the density fluctuations and phase coherence of the atomic field as functions of the system temperature.  We identify the onset of Bose condensation through analysis of both the generalized Binder cumulant appropriate to the inhomogeneous system, and the suppression of the effective many-body $T$ matrix that characterizes interactions between condensate atoms in the finite-temperature field.  We find that the system undergoes a second-order transition to condensation near the critical temperature for an ideal Bose gas in the strongly anisotropic three-dimensional geometry, but remains in a strongly phase-fluctuating quasicondensate regime until significantly lower temperatures.  We characterize the crossover from a quasicondensate to a true condensate by a qualitative change in the form of the non-local first-order coherence function of the field, and compare our results to those of previous works employing a density-phase Bogoliubov--de~Gennes analysis.      
\end{abstract}


\date{\today}

\maketitle

\section{Introduction}
Advances in the experimental control, observation and manipulation of quantum-degenerate dilute atomic gases have led to a large body of work focusing on the role of geometry and dimensionality in the physics of these quantum fluids \cite{Greiner2001a,Moritz2003a,Stock2005a,Trebbia2006a,Kinoshita2006a,vanAmerongen2008a,Clade2009a,Hung2011a,Jacqmin2011a,Yefsah2011a,Davis2012a}.  In homogeneous systems, Bose-Einstein condensation (BEC) of an ideal gas is precluded in dimensions $d<3$, and long-range order in finite-temperature interacting systems is prohibited in such low dimensionalities~\cite{Hohenberg1967a}.  Although some rigorous results exist for the harmonic trapping geometries typical of experimental dilute-gas systems~\cite{Mullin1997a,Lieb2002a,Fischer2002a,Fischer2005a}, these systems are theoretically less well characterized than their homogeneous counterparts.  Experimentally, these systems may exhibit \emph{quasicondensate} behavior, characterized by large phase fluctuations and comparatively subdued density fluctuations, similar to that predicted to occur in homogeneous low-dimensional systems~\cite{Popov1983,Kagan1987a}.  On the other hand, the finite size of such systems can induce some phase coherence across the spatial extent of the atomic sample.  This might loosely be associated with ``finite-size'' condensation, which will not exhibit the \emph{extensivity} property of a formal, thermodynamic Bose condensate~\cite{Penrose1956a}.  The physics of finite-sized and inhomogeneous quantum fluids in low dimensions are often more subtle than those of the infinite homogeneous systems for which more rigorous results are known (see, for example, Refs.~\cite{Kagan2000,Hadzibabic2009}), and the effects of finite-size condensation and its relationship to quasicondensation and superfluidity in low-dimensional systems remain in general somewhat unclear~\cite{Simula2008a,Bisset2009,Holzmann2008a,Desbuquois2012a}.

\newpage
A low-dimensional system of particular interest is the weakly interacting Bose gas in an elongated (cigar-shaped) three-dimensional (3D) harmonic trap~\cite{Dettmer2001a,Greiner2001a,Shvarchuck2002a,Richard2003a,Moritz2003a,Hellweg2003a,Hugbart2005a,Trebbia2006a,Esteve2006a,Kinoshita2006a,vanAmerongen2008a,Jacqmin2011a,Davis2012a}.  In the limit of an extreme trap anisotropy, such that the oscillator energy spacing in the two tightly confined (transverse) dimensions is much larger than the energy scales associated with interactions and thermal fluctuations, this system can be regarded as one-dimensional~\cite{Olshanii1998,Ho1999,Petrov2000a,Cazalilla2011a}.  Using a Bogoliubov--de~Gennes (BdG) approach in density and phase fluctuations, Petrov \emph{et al.}~\cite{Petrov2000a} determined the phase diagram for this system, and identified a regime of quasicondensate behavior.  Their analysis nevertheless showed that, at sufficiently low temperatures, long-wavelength fluctuations of the phase are suppressed.  The system is then phase coherent across a large spatial extent, and is said to contain a ``true'' condensate.   

Importantly, Bose gases in less severely elongated harmonic traps, in which the energy scales associated with interactions and thermal fluctuations are \emph{not} smaller than the transverse oscillator spacings (and which are therefore formally 3D), can also exhibit quasicondensate behavior.   This was demonstrated theoretically by Petrov~\emph{et al.}~\cite{Petrov2001a}, who adapted their BdG density-phase approach to the elongated 3D geometry.  Using an analytic hydrodynamic approximation for the structure of the axial BdG eigenfunctions, and assuming classical (equipartition) occupation numbers for these modes, they calculated the amplitude of the phase fluctuations, and hence the phase coherence length $l_\phi$, as functions of temperature.  As in the purely one-dimensional case, phase fluctuations in this system become suppressed at low temperatures, yielding a gradual crossover from a quasicondensate regime to a true condensate.  The authors of Ref.~\cite{Petrov2001a} characterized the crossover to the true-condensate regime by identifying a temperature $T_\phi$ below which the phase coherence length $l_\phi$ is larger than the extent of the quasicondensate.  The pronounced effect that interactions have on the behavior of this system is seemingly in stark contrast to the familiar case of unambiguously three-dimensional harmonically trapped Bose gases, in which interactions serve only to slightly decrease the critical temperature for BEC (see, e.g., Ref.~\cite{Davis2006a}).  The predicted quasicondensate behavior of the elongated 3D Bose gas has since been observed experimentally by a number of groups~\cite{Dettmer2001a,Richard2003a,Hellweg2003a,Hugbart2005a,Esteve2006a}.   

Although yielding great insight into the physics of phase-fluctuating condensates in 3D, the approach of Petrov and co-workers~\cite{Petrov2000a,Petrov2001a,Petrov2004} is only approximate in nature.  It neglects the effects of density fluctuations and their coupling to the phase fluctuations, and does not include the effects of fluctuations in the transverse dimensions.  As the density-phase decomposition on which it is based assumes the existence of a quasicondensate with suppressed density fluctuations, it cannot describe the gradual extinction of the quasicondensate, and the return of the system to the normal phase, with increasing temperature.  Moreover, as the fluctuations of the field are described without reference to an underlying condensate, the nature of the transition to condensation is obscured in such an approach.  It is therefore not clear from such calculations to what extent the familiar picture of BEC as a second-order transition remains relevant to the physics of the elongated case.  A thorough understanding of the relationship between condensate and quasicondensate in this comparatively straightforward scenario would seem to be a natural starting point for understanding the role of (potentially strictly finite-size) condensation in low-dimensional systems. 

In this article we apply the well-developed machinery of classical-field methods~\cite{Blakie2008a,Wright2011a} to understand the emergence of condensation and quasicondensation in a weakly interacting Bose gas in elongated 3D harmonic confinement.  The only limitation to this approach is the classical-field approximation itself; i.e., the neglect of the effects of quantum fluctuations, which are significant only at low temperatures.  We make use of a grand-canonical variant of the classical-field method~\cite{Gardiner2003a,Blakie2008a,Davis2013b} that is fully three-dimensional and includes the effects of interactions nonperturbatively~\cite{Blakie2008a}.  This method allows us to carefully characterize the condensate and quasicondensate, and the relationship between the two.  Although related methods have been used in several studies of the condensate--quasicondensate crossover in (quasi-)one-dimensional systems~\cite{AlKhawaja2002a,AlKhawaja2002aE,Proukakis2006b,Cockburn2011a,Bienias2011a,Gallucci2012a}, the only prior classical-field investigation of the elongated 3D system is that of Kadio \emph{et al.}~\cite{Kadio2005a}.  The authors of Ref.~\cite{Kadio2005a} calculated the phase-coherence length of the elongated 3D system, and found approximate agreement with the predictions of Petrov~\emph{et al.}~\cite{Petrov2001a}.  However, their study employed somewhat \emph{ad hoc} techniques --- based on ideal-gas arguments --- to estimate the temperature of the strongly fluctuating equilibrium state of the field, and was not able to access the statistics of the condensate mode itself, due to technical limitations~\cite{Kadio2005a}.  As such, no detailed exposition of the relation between condensation and quasicondensation in the elongated 3D system is available in the previously published literature.   

Here we present an extensive, quantitative analysis of the progression from normal gas to quasicondensate to true condensate in the elongated 3D gas with decreasing temperature.  We focus in particular on the nature of condensation in the system, which we associate with the orbital corresponding to the largest eigenvalue of the one-body density matrix~\cite{Penrose1956a}.  We characterize the onset of condensation by examining the number fluctuations of this condensate orbital, and the \emph{anomalous} correlations of the part of the field orthogonal to the condensate, and thereby show that the gas exhibits meaningful Bose condensation in the quasicondensate regime.  Moreover, we find that the onset of quasicondensation in the system is accompanied by a minimum in the effective interaction strength between condensate atoms that is consistent with the system undergoing a second-order transition to condensation.  We find that this occurs at a temperature slightly below the transition temperature of the corresponding ideal-gas model, and that the appearance of a quasicondensate in the system is therefore associated with the condensation transition.  The crossover from a quasicondensate to a true condensate at lower temperatures can thus be understood in terms of the correlations of the complementary \emph{noncondensed} component of the field, which we find are inconsistent with well-defined quasiparticle excitations --- i.e., with Gaussian, or Hartree-Fock-Bogoliubov (HFB) correlations~\cite{Blaizot86,Griffin96} --- at high temperatures, but come to be more consistent with HFB correlations as the system temperature is reduced.    

This article is organized as follows: In Sec.~\ref{sec:theory} we describe the theoretical methods used in our analysis.  We briefly explain our classical-field approach and the stochastic projected Gross-Pitaevskii equation (SPGPE) with which we describe the low-energy region of the system (Sec.~\ref{subsec:SPGPE}), and discuss how we use it to calculate observables of interest (Sec.~\ref{sec:observables}).  In Sec.~\ref{sec:results} we define the physical parameters of the system we investigate (Sec.~\ref{sec:params}) and present our analysis of its physical properties at varying temperature and fixed total atom number (Secs.~\ref{sec:condquasicond} -- \ref{sec:crosspt}).  In Sec.~\ref{sec:concl} we summarize our results and present our conclusions.

\section{Theoretical methods} \label{sec:theory}
\subsection{Stochastic projected Gross-Pitaevskii equation}\label{subsec:SPGPE}
The stochastic projected Gross-Pitaevskii equation (SPGPE) method, developed in Refs.~\cite{Gardiner2002a,Gardiner2003a,Bradley2008a}, has been reviewed in detail, together with other projected classical-field methods, in Ref.~\cite{Blakie2008a} (see also Refs.~\cite{Davis2013b,Wright2013a}).  For the reader's convenience, we briefly describe the relevant details of the formalism here.

Formally, the physics of the harmonically trapped dilute Bose gas is governed by the second-quantized Hamiltonian
\begin{align}\label{eq:H_fundamental}
    \hat{H} = \int & \!d\rbf\,\hat{\Psi}^\dagger(\rbf)H_\mathrm{sp}\hat{\Psi}(\rbf) \\ \nonumber
       & \!\! + \frac{1}{2}\int\!d\rbf\!\int\!d\rbf'\,\hat{\Psi}^\dagger(\rbf)\hat{\Psi}^\dagger(\rbf') U(\rbf-\rbf')\hat{\Psi}(\rbf')\hat{\Psi}(\rbf), 
\end{align}
where the single-particle Hamiltonian is
\begin{equation}\label{eq:Hsp}
	H_\mathrm{sp} = \frac{-\hbar^2\nabla^2}{2m} + \frac{m}{2}\Big[\omega_x^2x^2 + \omega_y^2y^2 + \omega_z^2z^2\Big],
\end{equation}
and $U(\rbf)$ is the exact interatomic potential.  We introduce a single-particle subspace $\mathbf{L}$ spanned by eigenmodes $Y_n(\rbf)$ of the single-particle Hamiltonian [$H_\mathrm{sp}Y_n(\rbf)=\epsilon_nY_n(\rbf)$] with energies $\epsilon_n$ less than a cutoff energy $E_\mathrm{max}$, and a complementary subspace composed of the remaining high-energy modes.  Provided $E_\mathrm{max}$ is chosen such that the high-energy modes are essentially unoccupied, the dynamics of these modes can be integrated out to obtain an effective Hamiltonian for the low-energy (coarse-grained) Bose field $\hat{\Psi}_\mathbf{L}(\rbf)=\sum_{n\in\mathbf{L}}\hat{a}_nY_n(\rbf)$, as shown by Morgan~\cite{Morgan2000}.  Atomic interactions described by the effective Hamiltonian are mediated by an approximate two-body $T$~matrix, and the interaction can thus be rigorously approximated by a ``contact" potential, with a renormalized coupling constant $U_0$.  In practice the correction due to the finite momentum cutoff is small~\cite{Norrie2006a}, and so we assume the standard $s$-wave coupling constant $U_0=4\pi\hbar^2a/m$, with $a$ the $s$-wave scattering length.  The low-energy Hamiltonian then takes the form 
\begin{align}\label{eq:H_L}
    \hat{H}_\mathrm{L} = \int & \!d\rbf\,\hat{\Psi}_\mathbf{L}^\dagger(\rbf)H_\mathrm{sp}\hat{\Psi}_\mathbf{L}(\rbf)  \nonumber \\
    & \!\! + \frac{U_0}{2}\!\int\!d\rbf\,\hat{\Psi}_\mathbf{L}^\dagger(\rbf)\hat{\Psi}_\mathbf{L}^\dagger(\rbf)\hat{\Psi}_\mathbf{L}(\rbf)\hat{\Psi}_\mathbf{L}(\rbf),  
\end{align}
which defines an effective field theory~\cite{Andersen2004} for the coarse-grained field $\hat{\Psi}_\mathbf{L}(\rbf)$.

We then further divide the low-energy region $\mathbf{L}$ into a coherent region (or condensate band) $\mathbf{C}=\{n : \epsilon_n < \epsilon_\mathrm{cut}\}$, spanned by single-particle eigenmodes $Y_\mathrm{n}(\rbf)$ with energies below some classical-field cutoff $\epsilon_\mathrm{cut}$ (the choice of which is discussed in Appendix~\ref{apx:cutoff}), and a complementary incoherent region $\mathbf{I}=\{n : \epsilon_\mathrm{cut} \leq \epsilon_n < E_\mathrm{max}\}$.  Introducing the projector 
\begin{equation}\label{eq:Pdef} 
    \mathcal{P}\C\big\{f(\rbf)\big\}\equiv\sum_{n \in  \mathbf{C}}Y_n(\rbf)\!\int\!d\rbf'\, Y_n^*(\rbf')f(\rbf'),  
\end{equation} 
onto the coherent region $\mathbf{C}$, we define a $\mathbf{C}$-region field operator
\begin{equation}\label{eq:C_region_opr}
    \hat{\psi}\C(\rbf) \equiv  \mathcal{P}\C\big\{ \hat{\Psi}_\mathbf{L}(\rbf) \big\} = \sum_{n\in\mathbf{C}} \hat{a}_n Y_n(\rbf).
\end{equation}
In the SPGPE formalism, the $\mathbf{C}$-region field operator $\hat{\psi}\C(\rbf)$ is treated in an open-systems approach, and the complementary $\mathbf{I}$ region of the field is regarded as a thermal and diffusive bath to which the $\mathbf{C}$ region is coupled.  The resulting master equation for the $\mathbf{C}$-region density operator corresponding to $\hat{\psi}\C(\rbf)$ is simplified by a high-temperature approximation and (after neglecting terms which do not affect the \emph{equilibrium} properties of the system~\cite{Bradley2008a,Rooney2012}) is mapped, using standard techniques~\cite{Gardiner2000}, onto a stochastic field equation in the Wigner representation~\cite{Steel1998, Sinatra2001} for a classical field 
\begin{equation}
	\psi\C(\rbf,t) = \sum_{n\in\mathbf{C}}\alpha_n(t)Y_n(\rbf).
\end{equation}
The resulting equation of motion 
\begin{align}
d\psi\C(\rbf,t) = \mathcal{P}\C & \bigg\{ \! -\frac{i}{\hbar} \mathcal{L}\C\psi\C(\rbf,t)dt \label{eq:SPGPE} \\
					 & \!\! + \frac{\gamma}{\kB T}
\left[\mu - \mathcal{L}\C\right]\psi\C(\rbf,t)dt + dW_{\gamma}(\rbf,t) \bigg\}, \nonumber
\end{align}
is termed the \emph{simple growth} SPGPE \cite{Bradley2008a,Blakie2008a}.
  
The growth rate $\gamma$ quantifies the strength of thermal and diffusive damping of the $\mathbf{C}$-region field $\psi\C(\rbf,t)$ by the high-energy bath of atoms in $\mathbf{I}$, and $dW_{\gamma}(\rbf,t)$ is a complex stochastic noise term associated with this damping, which satisfies
\begin{equation} \label{eq:noiseavg}
\langle dW^{*}_{\gamma}(\rbf,t)dW_{\gamma}(\rbf',t) \rangle = 2\gamma \delta\C(\rbf,\rbf')dt,
\end{equation}
where $\delta\C(\rbf,\rbf')=\sum_{n \in  \mathbf{C}}Y_n(\mathbf{r})Y_n^*(\mathbf{r}')$ acts as a Dirac delta function within the $\mathbf{C}$ region. The Hamiltonian evolution operator $\mathcal{L}\C$ for the $\mathbf{C}$ region is defined by its action on the $\mathbf{C}$-region field:
\begin{equation}
\mathcal{L}\C\psi\C(\rbf,t) \equiv \bigg( \! H_\mathrm{sp} + U_0\abs{\psi\C(\rbf,t)}^2 \! \bigg) \psi\C(\rbf,t).
\end{equation}
Neglecting all but the first term on the right-hand side (RHS) of Eq.~\reff{eq:SPGPE} we obtain the projected Gross-Pitaevskii equation (PGPE) \cite{Davis2001b,Davis2001a,Blakie2005a}.  The second term on the RHS of Eq.~\reff{eq:SPGPE} is dissipative and, in general, induces changes in the population $N\C=\int\!d\rbf\,|\psi\C(\rbf)|^2$ and energy $E\C = \int\!d\rbf\,\psi\C^*(\rbf)[H_\mathrm{sp}+(U_0/2)|\psi\C(\rbf)|^2]\psi\C(\rbf)$ of the $\mathbf{C}$-region field.  Within this term, $\mathcal{L}\C$ can be thought of as extracting the effective (local) chemical potential of the classical field; neglecting for simplicity the phase of the field, the local field amplitude $\psi\C(\rbf)$ therefore grows where $\mathcal{L}\C\psi\C(\rbf)$ is smaller than $\mu\psi\C(\rbf)$, and \emph{vice versa}.  The complex noise term $dW_\gamma(\rbf,t)$ reflects the stochastic nature of the dissipation, which results physically from the random scattering of atoms into and out of the $\mathbf{C}$ region.  

An expression for the growth rate $\gamma$ in terms of the thermodynamic parameters of the bath and the choice of energy cutoff $\epsilon_\mathrm{cut}$ was derived systematically in Ref.~\cite{Bradley2008a}.  However, in the present study the precise value of $\gamma$ is unimportant, as we are only concerned with the equilibrium properties of the system, and not the detailed nonequilibrium dynamics of its relaxation.  We thus choose a value for $\gamma$ on the basis of numerical expediency (see Sec.~\ref{sec:params}).  

\subsection{Calculation of observables} \label{sec:observables}
The noise and damping terms in the SPGPE [Eq.~\reff{eq:SPGPE}] serve to drive trajectories of the classical field to a grand-canonical equilibrium distribution consistent with the imposed (thermal bath) temperature~$T$ and chemical potential~$\mu$~\cite{Blakie2008a}.  The field undergoes a period of non-equilibrium dynamical evolution as it thermalizes toward equilibrium with the (above-cutoff) $\mathbf{I}$ region.  The growth of the (quasi-)condensate from an evaporatively cooled thermal cloud would, in principle, be modeled by starting with an initial state $\psi\C(\rbf,t=0)$ corresponding to a high-temperature, noncondensed field~\cite{Bradley2008a,Weiler2008a}.  However, as our interest here is in the equilibrium configurations of the system, we choose for our initial state the ground state of the Gross-Pitaevskii equation, which we obtain for each considered chemical potential $\mu$ by imaginary-time evolution.  In this way we avoid the spontaneous formation of long-lived phase defects during the passage of the system from the noncondensed to the \mbox{(quasi-)}condensed phase \cite{Zurek1996,Anglin1999,Weiler2008a}, which can significantly delay complete thermalization~\cite{Berloff2002}.  

Once equilibrium is established, we characterize the state of the system by calculating correlation functions of the classical field; i.e., averages of functionals $\mathcal{F}[\psi\C(\rbf)]$ of the field over the equilibrium distribution of field configurations~\cite{Blakie2005a}.  The $\mathbf{I}$ region itself is modeled in a semiclassical Hartree-Fock approximation~\cite{Davis2006a,Rooney2010} (see Appendix~\ref{apx:SCHF}), from which we can infer the total field density $n(\rbf) = n\C(\rbf) + n\I(\rbf)$, and the total atom number $N=\int\! d\mathbf{r}\, n(\rbf) \equiv N\C+N\I$.  The classical correlation functions of $\psi\C(\rbf)$ are the classical-field analogs of quantum correlation functions of the Bose field $\hat{\psi}\C(\rbf)$~\cite{Blakie2005a}, and we interpret them as estimates of the corresponding quantum correlation functions (i.e., we neglect the formal commutator corrections of the Wigner theory~\cite{Blakie2008a}, which is equivalent to neglecting quantum fluctuations).  In practice, we substitute time averages of a single trajectory $\psi\C(\rbf,t)$ for averages over the grand-canonical ensemble: 
\begin{equation}
    \langle \mathcal{F}[\psi\C(\rbf)] \rangle = \frac{1}{N_s}\sum_{j=1}^{N_s} \mathcal{F}[\psi\C(\rbf,t_j)],
\end{equation}
following the ergodic interpretation of the formally microcanonical classical-field methods~\cite{Davis2002a,Goral2002a,Davis2005a}.

\subsubsection{Coherence and (quasi-)condensation}
In this article we consider only equal-time correlations of the classical field $\psi\C$.  The expectation values of all one-body observables in the field at equilibrium are encoded by the first-order coherence function~\cite{Mandel1995,Naraschewski1999a,NoteA}
\begin{equation}
	G^{(1)}(\rbf,\rbf') = \langle \psi\C^{*}(\rbf)\psi\C(\rbf') \rangle.
\end{equation}
The local ($\rbf'=\rbf$) first-order coherence function yields the mean density of atoms in the $\mathbf{C}$ region, $n\C(\rbf) = G^{(1)}(\rbf,\rbf)$,  while the off-diagonal ($\rbf'\neq\rbf$) elements $G^{(1)}(\rbf,\rbf')$ depend additionally on the coherence of the phase $\phi(\rbf)$ of the classical field [defined by $\psi\C(\rbf)=|\psi\C(\rbf)|e^{i\phi(\rbf)}$] between positions $\rbf$ and $\rbf'$.   The matrix $G^{(1)}$ is Hermitian, and can therefore be diagonalized to obtain a complete basis of eigenvectors with real eigenvalues.  Transposing the Penrose-Onsager definition~\cite{Penrose1956a} of BEC to the classical-field description, we identify the largest of these eigenvalues as the condensate population $N_0$, and the corresponding eigenvector $\varphi_0(\rbf)$ as the (unit-normalized) condensate orbital.  This identification is supported \emph{ex post facto} by a consideration of higher-order field correlations, as we discuss in Sec.~\ref{sec:results}.

The inhomogeneous system we investigate may exhibit more general \emph{quasicondensate} behavior, associated with the slow decay of phase coherence across the sample~\cite{Petrov2000a,Petrov2001a,Petrov2004}.  We will therefore make use of the normalized first-order coherence function
\begin{equation}
	g^{(1)}(\rbf,\rbf') = \frac{G^{(1)}(\rbf,\rbf')}{\sqrt{n\C(\rbf)n\C(\rbf')}},
\end{equation}
to characterize the spatial decay of phase coherence in the system.

Another important characterization of the field fluctuations is given by the normalized local second-order coherence function
\begin{equation}
	g^{(2)}(\rbf,\rbf) = \frac{\langle |\psi\C(\rbf)|^4 \rangle }{\langle |\psi\C(\rbf)|^2\rangle^2},
\end{equation}
which is directly related to the density variance: $\mathrm{Var}\{n\C(\rbf)\} = (g^{(2)}(\rbf,\rbf) - 1) {n\C}^2(\rbf)$~\cite{NoteB}, and therefore provides a measure of density fluctuations. In the limiting case of a purely thermal (chaotic) field, $g^{(2)}(\rbf,\rbf)=2$, whereas $g^{(2)}(\rbf,\rbf)=1$ for a perfectly coherent field~\cite{Blakie2005a,Glauber1965,Mandel1995}.   

In a homogeneous system, a quasicondensate is a component of the field that undergoes large point-to-point phase fluctuations, but in which density fluctuations are suppressed~\cite{Popov1983,Kagan1987a}.  The quasicondensate density can therefore be estimated by considering the extent to which fluctuations of the field fail to be Gaussian~\cite{Prokofev2001a,Prokofev2002}.  Although such identifications in general bear no \emph{a priori} relation to the structure of nonlocal phase correlations in the field, they do provide a useful characterization of the quasicondensate~\cite{Prokofev2001a,Prokofev2002}.  Generalizing to the inhomogeneous case, the quantity~\cite{NoteA} 
\begin{equation}
	n\QC(\rbf) = \sqrt{2{n\C}^2(\rbf) - G^{(2)}(\rbf,\rbf)},
\end{equation}
provides a useful measure of the field statistics in experimentally relevant systems~\cite{Proukakis2006b,Bisset2009}, and is often simply referred to as the \emph{quasicondensate density} --- a terminology we will also adopt in this article.  Following from this definition we define the total quasicondensate population $N\QC \equiv \int\!d\rbf\, n\QC(\rbf)$. 

\subsubsection{Anomalous correlations} \label{sec:anomalous}
We further characterize the physics of our system by analyzing the so-called anomalous correlations of the field fluctuations.  In contrast to traditional mean-field theories of BEC (see, e.g., Ref.~\cite{Griffin96}), our approach does not assume \emph{a priori} a particular fixed value for the condensate phase, and hence preserves $\mathrm{U(1)}$ phase symmetry, in the sense that the grand-canonical ensemble sampled by the classical field is invariant under global phase rotations.  We therefore follow Ref.~\cite{Wright2011a} (see also Refs.~\cite{Wright2010a,Cockburn2011a,Sinatra2011a,Vladimirova2011a}) in defining the \emph{fluctuation field}  
\begin{equation} \label{eq:Lambda}
	\Lambda(\rbf,t)\equiv \frac{\alpha_0^*(t)}{\sqrt{\alpha_0^*(t)\alpha_0(t)}}\delta\psi\C(\rbf,t),
\end{equation}
where $\alpha_0(t)=\int\!d\rbf\,\varphi^*_0(\rbf)\psi\C(\rbf,t)$ is the classical-field amplitude of the (unit-normalized) condensate mode $\varphi_0$~\cite{NoteC}, and
\begin{equation}
	\delta\psi\C(\rbf,t)=\psi\C(\rbf,t) - \alpha_0(t)\varphi_0(\rbf),
\end{equation}
is the component of the classical field $\psi\C(\rbf,t)$ orthogonal to the condensate.  Introducing $\Lambda(\rbf)$ allows us to calculate anomalous moments of the non-condensed component of the field, and we will consider in particular the so-called anomalous thermal density (or anomalous average) $\kappa(\rbf) = \langle \Lambda(\rbf)\Lambda(\rbf) \rangle$~\cite{NoteA}.  Whereas the quantity $\langle \delta\psi\C(\rbf)\delta\psi\C(\rbf) \rangle$ vanishes in the grand-canonical ensemble sampled by the SPGPE, which is symmetric with respect to global phase rotations, $\kappa(\rbf)$ can acquire a nonzero value in this ensemble, giving a measure of ``pairing'' correlations induced in the noncondensed component of the field by the condensate~\cite{Blaizot86}.  Physically, these correlations arise due to the coherent scattering of pairs of atoms out of the condensate~\cite{Morgan2000}, and the corresponding time-reversed processes.  At equilibrium the rates of forward and reverse scattering must balance, and so we expect $\kappa(\rbf)$ to be purely real, relative to a real condensate orbital~\cite{Wright2011a,Vladimirova2011a}.

\section{Results} \label{sec:results}
\subsection{Physical parameters} \label{sec:params}
We consider a finite-temperature cloud of $^{23}$Na atoms in an elongated (cigar-shaped) harmonic potential with longitudinal and transverse trapping frequencies $\omega_z = 2\pi\times5$~Hz and $\omega_x=\omega_y\equiv\omega_\perp=2\pi\times250$~Hz, respectively.  Results for some physical quantities will be specified in terms of the (long-axis) oscillator length $z_0=\sqrt{\hbar/m\omega_z}$ and oscillator energy $\hbar\omega_z$.  We focus primarily on the dependence of the field correlations on the system temperature at constant \emph{total} ($\mathbf{C}$-region plus $\mathbf{I}$-region) atom number $N=2\times10^5$.  

As the system we study is fundamentally three-dimensional, we expect the transition temperature of the corresponding ideal Bose gas to provide a useful point of comparison (in contrast to inherently one- and two-dimensional systems, in which interactions can easily render the corresponding ideal-gas models irrelevant~\cite{Bouchoule2007a,Holzmann07}).  We calculate the critical temperature for the Bose gas in the elongated potential in the non-interacting limit 
\begin{equation}\label{eq:Tc0}
	\Tc^0 \approx \frac{\hbar\bar{\omega}}{\kB} \left[\left(\frac{N}{\zeta(3)}\right)^{\frac{1}{3}} - \frac{1}{2}\frac{\zeta(2)}{\zeta(3)}\frac{\omega_m}{\bar{\omega}}\right],
\end{equation}
where $\zeta$ is the Riemann zeta function, $\bar{\omega} \equiv (\omega_x\omega_y\omega_z)^{1/3}$, and $\omega_m \equiv (\omega_x + \omega_y + \omega_z)/3$. The second term on the RHS of Eq.~\reff{eq:Tc0} is the (leading-order) finite-size correction to the critical temperature due to the trap anisotropy \cite{Grossmann1995a,Pethick2008a}.  For our system Eq.~\reff{eq:Tc0} gives a transition temperature $T_c^0\approx 174\;\nK$.  At temperatures below $\Tc^0$, the condensate fraction of the ideal gas varies as $N_0/N \approx 1 - (T/T_c^0)^3$~\cite{Pethick2008a}. 

We simulate the field at a range of temperatures $T = 85$ -- $180\;\nK$.  This range extends from the lowest temperature at which the validity criteria of the SPGPE can be satisfied to just above the ideal-gas estimate of the BEC transition temperature [Eq.~\reff{eq:Tc0}].  We discuss the rationale behind our choice of system parameters further in Appendix~\ref{apx:cutoff}.  Characterizations of the system at constant chemical potential $\mu$, and at constant temperature $T$, are reported in Appendix~\ref{apx:fixed_mu_and_fixed_T}.

As our classical-field method is grand canonical, in order to perform calculations at fixed total atom number $N$, we must first calculate the appropriate chemical potential $\mu(N,T)$.  This is achieved using a semiclassical Hartree-Fock description of the above-cutoff atoms (Appendix~\ref{apx:SCHF}).  For simulations performed at fixed $\mu$ or fixed $T$ (Appendix~\ref{apx:fixed_mu_and_fixed_T}), we simply calculate the above-cutoff density and total atom number after performing our classical-field simulations.  We choose values of the growth rate $\gamma$ corresponding to a fixed dimensionless growth coefficient $\gamma/\kB T = 40ma^2\kB T /\pi \hbar^3$ in all our simulations (cf. Refs.~\cite{Bradley2008a,Blakie2008a}).  We find for our choice of parameters that thermalization in each simulation occurs over a time scale of approximately $100\;\ms$ of system time, though to ensure sufficient equilibration we form our time averages from $10^4$ equally spaced samples taken over a period of $100$~s, beginning after the first $1$~s of evolution.    

\subsection{Condensate and quasicondensate} \label{sec:condquasicond}
\begin{figure}
	\begin{center}
	\includegraphics{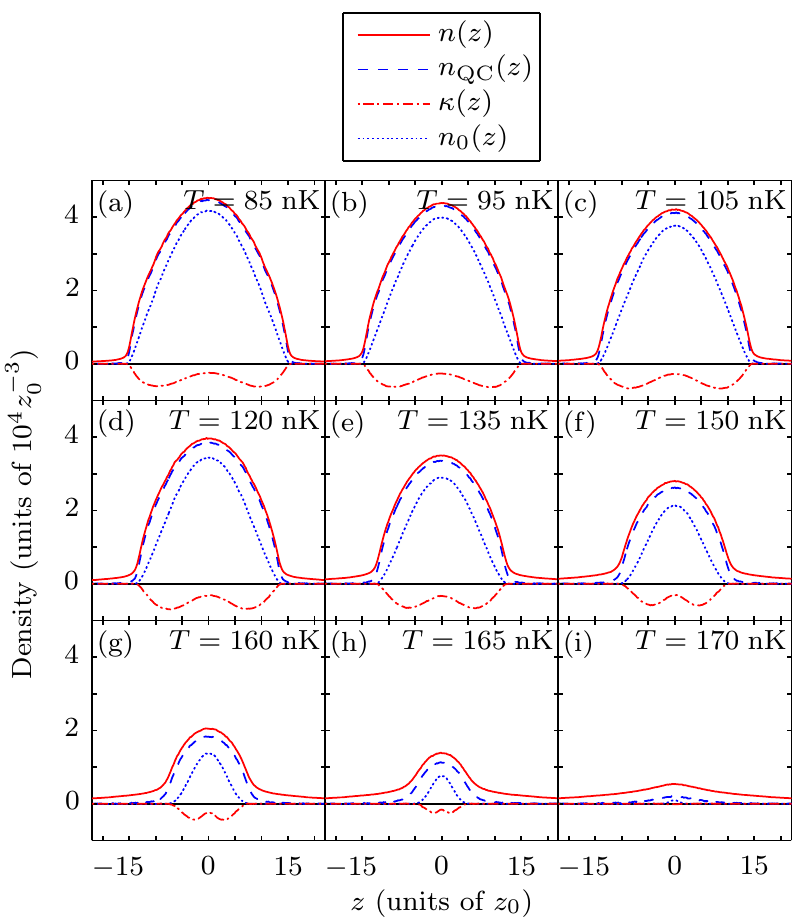}
	\end{center}
	\caption{\small (Color online) Variation of system densities on the long ($z$) axis with temperature: condensate density $n_0(z)$ (dotted blue line), quasicondensate density $n\QC(z)$ (dashed blue line), anomalous density $\kappa(z)$ (dot-dashed red line), and total ($\mathbf{C}$-region plus semiclassical $\mathbf{I}$-region) density $n(z)$ (solid red line).}
	\label{fig:anomalous}
\end{figure}

In Fig.~\ref{fig:anomalous} we show the condensate density $n_0(z)=N_0|\varphi_0(z\hat{\mathbf{z}})|^2$ (dotted blue line) and the quasicondensate density $n\QC(z)=n\QC(z\hat{\mathbf{z}})$ (dashed blue line), on the $z$~axis, at a set of representative temperatures.  We also plot the anomalous thermal density $\kappa(z)=\kappa(z\hat{\mathbf{z}})$ (dot-dashed red line), and the total density of the Bose field $n(z)=n\C(z\hat{\mathbf{z}})+n\I(z\hat{\mathbf{z}})$ (solid red line) [see Eq.~\reff{eq:nth}].  At the lowest temperatures considered [Figs.~\ref{fig:anomalous}(a)--\ref{fig:anomalous}(c)] we observe significant condensate density, and associated anomalous thermal density.  We find that the anomalous density is nonzero only in the central region of the trap where the condensate exists, and its magnitude exhibits a pronounced dip at the center of the trap, consistent with the results of previous works~\cite{Proukakis1998a,Hutchinson2000a,Bergeman2000,Wright2010a,Cockburn2011a,Wright2011a,Boudjemaa2011}.  

We note that even in this very low temperature regime, the quasicondensate density $n\QC(z)$ is somewhat larger than the condensate density $n_0(z)$~\cite{NoteD}.  However, this is also a feature of condensation in a less anisotropic (oblate) 3D trap, as observed in Ref.~\cite{Wright2011a}.  In particular, the small excess proportion of quasicondensate here is largely attributable to the presence of the anomalous average; i.e., it is mostly accounted for by the (negative) contribution of $(\varphi_0^*)^2\langle\Lambda\Lambda\rangle + \mathrm{c.c.}$ (not shown) to the local coherence function $G^{(2)}(\rbf,\rbf)$ (see Ref.~\cite{Wright2011a}).  In the limit that the system exhibits well-defined Bogoliubov quasiparticles, the anomalous average is a measure of the phase-fluctuation-like nature~\cite{Weichman1988a} of the lowest-lying excitations.   The phase-fluctuation character of these excitations implies that their thermal population contributes little to the density fluctuations of the sample, as compared (for example) to dressed single-particle (Hartree-Fock) states~\cite{Goldman1981a}.  This effect can be exhibited in standard mean-field calculations that linearize the field fluctuations about a well-defined condensate mode~\cite{Gardiner2001a}, and is implicit in the BdG density-phase approach of Refs.~\cite{Petrov2000a,Petrov2001a}.  In the sense that the field correlations at these low temperatures can be understood in terms of a distinct condensate with well-defined quasiparticle excitations, they are consistent with true condensation.   

As the temperature is increased [Figs.~\ref{fig:anomalous}(d)--\ref{fig:anomalous}(f)], the size of the quasicondensate progressively decreases, and the atomic density is increasingly redistributed to purely thermal population in the wings of the cloud.  At the same time, the condensate density recedes from the quasicondensate density and the anomalous density correspondingly becomes smaller.  At reasonably high temperatures [Figs.~\ref{fig:anomalous}(g)--\ref{fig:anomalous}(h)], meaningful condensation (as evidenced by the persistent presence of an associated anomalous average) remains, though both the condensate and anomalous average are significantly smaller than the quasicondensate.  In this regime, the large excess of quasicondensate density is no longer accounted for by the presence of the anomalous average.  In particular, the presence of significant quasicondensation outside the spatial extent of the condensate indicates that the quasicondensate is largely composed of higher-order correlations in the field~\cite{Wright2011a}, reflecting the breakdown of a description in terms of Bogoliubov quasiparticle excitations about a well-defined condensate~\cite{Fedichev1998b,Morgan2000}.  At these temperatures, the system is more appropriately characterized by a linearized expansion in phase and density fluctuations~\cite{Petrov2004}.   
Our SPGPE approach, which naturally accounts for all dynamical processes in the field (within the classical-field approximation), is equally applicable to the quasicondensate regime, the low-temperature true condensate regime, and the crossover between the two.  At the highest temperature shown [Fig.~\ref{fig:anomalous}(i)], a small quasicondensate remains, while the condensate (and the associated anomalous average) is vanishingly small.

We note that the progression of field correlations with increasing temperature shown here is qualitatively similar to that obtained in classical-field calculations for a more standard 3D geometry~\cite{Wright2011a}:  As the temperature of the field is increased, the condensate becomes an increasingly small proportion of the total quasicondensate, and the quasicondensate is increasingly composed of correlations beyond a simple Gaussian (Hartree-Fock-Bogoliubov~\cite{Blaizot86}) ansatz for field fluctuations about a well-defined condensate.  We therefore infer that the quasicondensate behavior of the present highly elongated sample has the same physical origin as the partial quasicondensation that arises already in a standard 3D geometry.  The elongation of the trap in the present scenario merely introduces a quantitative correction, causing the divergence of the condensate from the quasicondensate to occur at significantly lower temperatures, so that the quasicondensate-dominated regime extends over a larger temperature range~\cite{NoteE}. 

\begin{figure}
	\begin{center}
	\includegraphics{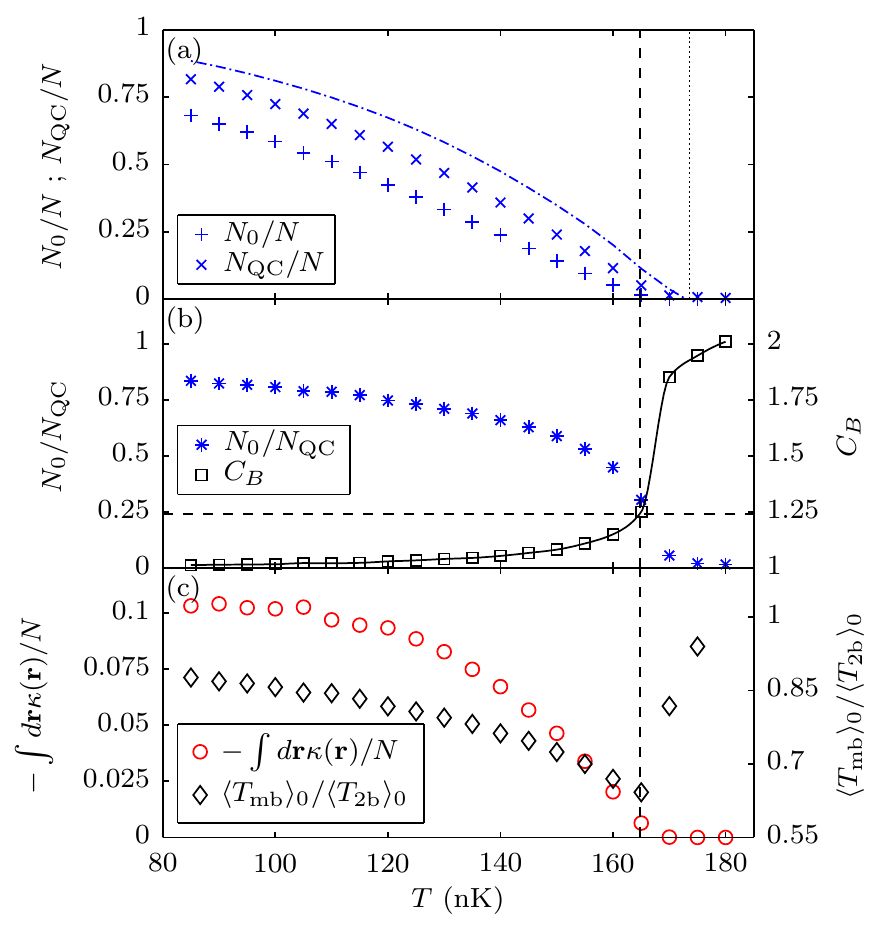}
	\end{center}
	\caption{\small (Color online) (a) Dependence of condensate (pluses) and quasicondensate (crosses) fractions on system temperature.  Also shown are the ideal-gas BEC transition temperature $T_c^0$ (vertical dotted line) and condensate fraction (dot-dashed line).  (b) Ratio of condensate and quasicondensate populations $N_0/N\QC$ (asterisks), and generalized Binder cumulant $C_B$ (squares).  The horizontal dashed line indicates the critical value $(C_B)_\mathrm{crit}$, and the vertical dashed line indicates the corresponding critical temperature.  The solid line is a smooth line of best fit used to identify the intersection with $(C_B)_\mathrm{crit}$.  (c) (Negative of the) integrated anomalous density (circles), and ratio of expectation values of the effective many-body $T$~matrix and two-body $T$~matrix in the condensate (diamonds).}
	\label{fig:Nconst}
\end{figure}

In Fig.~\ref{fig:Nconst}(a) we plot the condensate and quasicondensate populations ($N_0$ and $N\QC$, respectively) as fractions of the total atom number $N$.  The occupancies of the condensate and quasicondensate exhibit similar trends, smoothly increasing from near zero at high temperatures toward $N$ as the temperature decreases toward zero.  Interestingly, the functional dependence of $N_0$ on temperature appears to correspond more closely to the $N_0/N\sim 1 - T/T_c$ scaling of the condensate fraction of an ideal gas in one-dimensional harmonic confinement~\cite{Ketterle1996a}, than that of the full 3D geometry (Sec.~\ref{sec:params}).

The marked difference between the occupancies of the condensate and quasicondensate is illustrated by their ratio $N_0/N\QC$, plotted in Fig.~\ref{fig:Nconst}(b) (asterisks). This ratio is small at temperatures $T \gtrsim 160$~nK, and the system is therefore dominated by quasicondensate behavior in this regime. At lower temperatures this ratio steadily increases with decreasing temperature, indicating a gradual crossover from a quasicondensate to a true condensate. 

\subsection{Identification of the BEC critical point} \label{sec:critpt}
\subsubsection{Binder cumulant analysis}
In Fig.~\ref{fig:Nconst}(b) we also plot the \emph{generalized Binder cumulant} $C_B\equiv \langle|\alpha_0|^4\rangle/\langle|\alpha_0|^2\rangle^2$ (squares), where $\alpha_0(t)$ is the condensate amplitude (see Sec. \ref{sec:observables}). Bezett and Blakie~\cite{Bezett2009a} introduced this quantity as the natural generalization of the Binder cumulant \cite{Binder1981a} of a homogeneous system to the harmonically trapped case.  In homogeneous models within the 3D $XY$ universality class, the Binder cumulant acquires the universal value $(C_B)_{\mathrm{crit}}=1.243$ at the critical point associated with the transition to long-range order \cite{Campostrini2001a}.   This fact was used to identify the critical temperature in classical-field simulations of the 3D homogeneous Bose gas~\cite{Davis2003a}.  More recently, we found~\cite{Wright2011a} that an estimate of the critical point of a harmonically trapped Bose gas based on the condition $C_B = (C_B)_\mathrm{crit}$ was consistent with an independent estimation based on the suppression of condensate-condensate interactions due to many-body effects~\cite{Proukakis1998a}.  Here we take the condition $C_B = (C_B)_\mathrm{crit}$ as an estimate of the location of the Bose-condensation transition in the elongated 3D system, and thereby identify the critical temperature $T_c = 165\;\nK$.  Our analysis of the Binder cumulant therefore indicates a transition to Bose condensation at a temperature slightly below the appropriate ideal-gas BEC temperature, consistent with previous results for less anisotropic harmonically trapped 3D systems~\cite{Davis2006a,Smith2011a}.

\subsubsection{Many-body T-matrix analysis}
In Fig.~\ref{fig:Nconst}(c) we plot (the negative of) the integrated anomalous density $\int\!d\mathbf{r}\,\kappa(\mathbf{r})$ (circles), which reaches its maximum absolute value at intermediate temperatures, consistent with previous studies of finite-temperature condensates~\cite{Proukakis1998a,Hutchinson2000a,Bergeman2000,Wright2010a,Cockburn2011a,Wright2011a,Boudjemaa2011}.  As we have noted (Sec.~\ref{sec:observables}), the anomalous density arises due to so-called Bogoliubov processes, in which pairs of atoms scatter each other out of the condensate and into the noncondensed modes of the field (and \emph{vice versa}).  These processes yield a correction to the strength of condensate-condensate interactions, due to Bose-stimulated scattering of colliding condensate atoms through \emph{occupied} intermediate modes.  In mean-field theories, this correction is encoded in the replacement of the $s$-wave (contact-potential~\cite{NoteF}) two-body $T$~matrix $T_\mathrm{2b}(\mathbf{r},\mathbf{r}') = U_0 \delta(\rbf-\rbf')$ by an approximate many-body $T$~matrix~\cite{Proukakis1998a,Morgan2000,Hutchinson2000a,Bergeman2000,Olshanii2001,Pricoupenko2011,Kim2012}
\begin{equation}
    T_\mathrm{mb}(\rbf,\rbf') = U_0\left(1 + \frac{\kappa(\rbf)}{N_0\varphi_0^2(\rbf)}\right)\delta(\rbf-\rbf').
\end{equation}

To quantify the total correction to condensate-condensate scattering due to this many-body effect, we calculate the matrix element of $T_\mathrm{mb}$ in the condensate orbital
\begin{align}
    \langle T_\mathrm{mb} \rangle_0 &\equiv \langle \varphi_0 \varphi_0 | T_\mathrm{mb} | \varphi_0 \varphi_0 \rangle \nonumber \\
                                    &= U_0 \int d\rbf \left(1+\frac{\kappa(\rbf)}{N_0\varphi_0^2(\rbf)}\right) |\varphi_0(\rbf)|^4,
\end{align}
and compare it with the corresponding matrix element of the two-body $T$~matrix, $\langle T_\mathrm{2b} \rangle_0 = U_0 \int\! d\rbf\, |\varphi_0(\rbf)|^4$.  The ratio $\langle T_\mathrm{mb} \rangle_0/\langle T_\mathrm{2b} \rangle_0$ of the two matrix elements is plotted in Fig.~\ref{fig:Nconst}(c) (black diamonds).

As is well known, the effect of the many-body processes encoded in $T_\mathrm{mb}$ is to suppress the strength of scattering between condensate atoms:  In a homogeneous 3D system the effective interaction strength between condensate atoms vanishes at the BEC phase transition ~\cite{Bijlsma1996,Bijlsma1997,Shi1998}.  More generally, one finds in inhomogeneous geometries that the ratio $\langle T_\mathrm{mb} \rangle_0 / \langle T_\mathrm{2b} \rangle_0$ exhibits a minimum at the critical temperature for Bose condensation~\cite{Proukakis1998a,Hutchinson2000a,Bergeman2000,Wright2011a,NoteG}.  From Fig.~\ref{fig:Nconst}(c) we observe that the ratio of effective $T$-matrix elements indeed exhibits a minimum value close to the critical temperature $T_c$ estimated from the analysis of the Binder cumulant (Sec.~\ref{sec:critpt}).  This behavior strongly suggests that despite the phase-fluctuating (quasicondensate) nature of the system at temperatures $T \lesssim T_c$, the system does indeed exhibit a second-order phase transition to condensation, and moreover provides an independent validation of the Binder cumulant condition $C_B = (C_B)_\mathrm{crit}$ used to estimate the critical point.  

\subsection{Phase coherence} \label{sec:phasecoh}
\begin{figure}
	\begin{center}
	\includegraphics{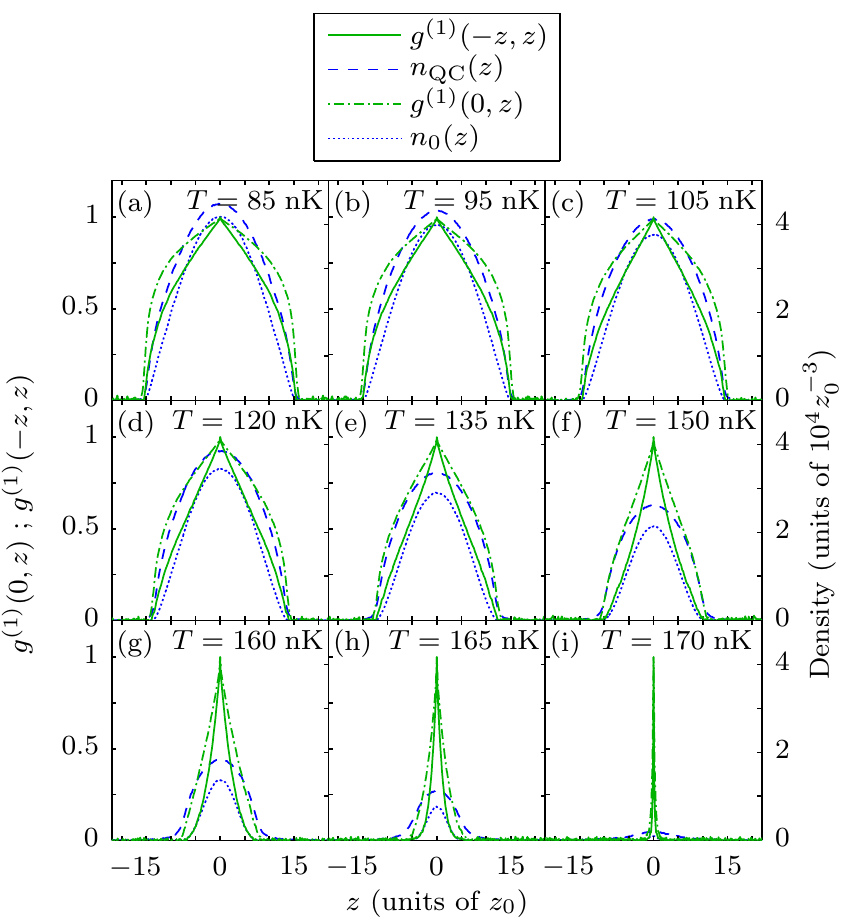}
	\end{center}
	\caption{\small (Color online) First-order correlation functions $g^{(1)}(-z,z)$ (solid green) and $g^{(1)}(0,z)$ (dot-dashed green), and densities of the condensate $n_0(z)$ (dotted blue) and quasicondensate $n\QC(z)$ (dashed blue).}
	\label{fig:g1density}
\end{figure}

To characterize the temperature dependence of the equilibrium behavior of the system in more detail, we analyze the normalized first-order correlation function on the $z$ (long) axis, $g^{(1)}(z,z')\equiv g^{(1)}(z \hat{\mathbf{z}}, z' \hat{\mathbf{z}})$, which reveals the spatial extent of phase coherence in the system.  In Fig.~\ref{fig:g1density} we plot the coherence function relative to the center of the system $g^{(1)}(0,z)$ (dot-dashed green line) and the symmetrically evaluated coherence function $g^{(1)}(-z,z)$ (solid green line), along with the condensate density $n_0(z)$ (dotted blue line) and quasicondensate density $n\QC(z)$ (dashed blue line), for the representative temperatures considered in Fig.~\ref{fig:anomalous}.  The condensate mode shape is of course determined by the form of $g^{(1)}(\rbf,\rbf')$ through its definition in terms of the Penrose-Onsager criterion.  By contrast, the density profile $n\QC(z)$ of the quasicondensate, which simply corresponds to the suppression of local density fluctuations, is \emph{a priori} unrelated to the first-order coherence of the system (see Sec.~\ref{sec:observables}).

Above the critical point [Fig.~\ref{fig:g1density}(i)], the spatial extent of both $g^{(1)}(0,z)$ and $g^{(1)}(-z,z)$ is very narrow, with both functions decaying on the length scale of the thermal de~Broglie wavelength $\lambda_T = \hbar\sqrt{2\pi/m\kB T}$, which in this regime is far smaller than the axial harmonic oscillator length $z_0=\sqrt{\hbar/m\omega_z}$ \cite{Naraschewski1999a}.  Correspondingly, the condensate itself is very small, whereas the quasicondensate is still finite across a comparatively large spatial extent, indicating a regime of density-fluctuation suppression without any significant one-body coherence.

At and slightly below the critical temperature [Figs.~\ref{fig:g1density}(h)~and~\ref{fig:g1density}(g), respectively] we observe markedly broader profiles for both $g^{(1)}(0,z)$ and $g^{(1)}(-z,z)$.  Moreover, the symmetrically evaluated correlation $g^{(1)}(-z,z)$ exhibits a roughly exponential decay with $z$, as expected in the quasicondensate regime~\cite{Petrov2001a,Gerbier2003a}.  The profiles of both these correlation functions remain far narrower than $n\QC(z)$, indicating that the system is well within the quasicondensate regime, consistent with the small magnitude of the condensate density $n_0(z)$ relative to that of the quasicondensate $n\QC(z)$.

At lower temperatures [Figs.~\ref{fig:g1density}(a)--\ref{fig:g1density}(f)] we see evidence of a crossover from a quasicondensate to a true condensate: The profiles of $g^{(1)}(0,z)$ and $g^{(1)}(-z,z)$ both broaden and become similar in width to the quasicondensate density $n\QC(z)$ as the temperature is decreased.  Furthermore, the shape of $g^{(1)}(-z,z)$ departs from exponential decay and comes to more closely resemble Gaussian decay at these lower temperatures~\cite{Gerbier2003a}, particularly in Figs.~\ref{fig:g1density}(a)--\ref{fig:g1density}(d), and the condensate density $n_0(z)$ approaches $n\QC(z)$ with decreasing temperature.

\subsection{Identification of the condensate-quasicondensate crossover temperature}\label{sec:crosspt}
To quantitatively characterize the crossover from quasicondensate to true condensate, we fit the function
\begin{equation}\label{eq:phase_fit}
	f(z) \equiv e^{-[(1-\xi)(2z/L_{\phi}) + \xi(2z/L_{\phi})^2]}
\end{equation}
to the correlation function $g^{(1)}(-z,z)$, where $L_\phi$ is the phase-coherence length, and $\xi \in [0,1]$ is a second fitting parameter that controls the functional form of $f(z)$.  The function $f(z)$ exhibits purely exponential decay in the limit $\xi=0$ --- characteristic of $g^{(1)}(-z,z)$ in the quasicondensate regime --- and purely Gaussian decay in the limit $\xi=1$ --- characteristic of $g^{(1)}(-z,z)$ in the true condensate regime~\cite{Gerbier2003a,Proukakis2006b}.  Using this fitting function has the advantage of taking into consideration both the width and shape of $g^{(1)}(-z,z)$ in providing an estimate of the coherence length $L_\phi$ \cite{Proukakis2006b}.  Moreover, it allows us to identify a \emph{qualitative} change in the correlation function~\cite{Gerbier2003a,Petrov2004} as a signature of the crossover, and in this article we define the characteristic temperature of the crossover as that at which $\xi=0.5$ (cf. Ref.~\cite{Proukakis2006b}).

\begin{figure}
	\begin{center}
	\includegraphics{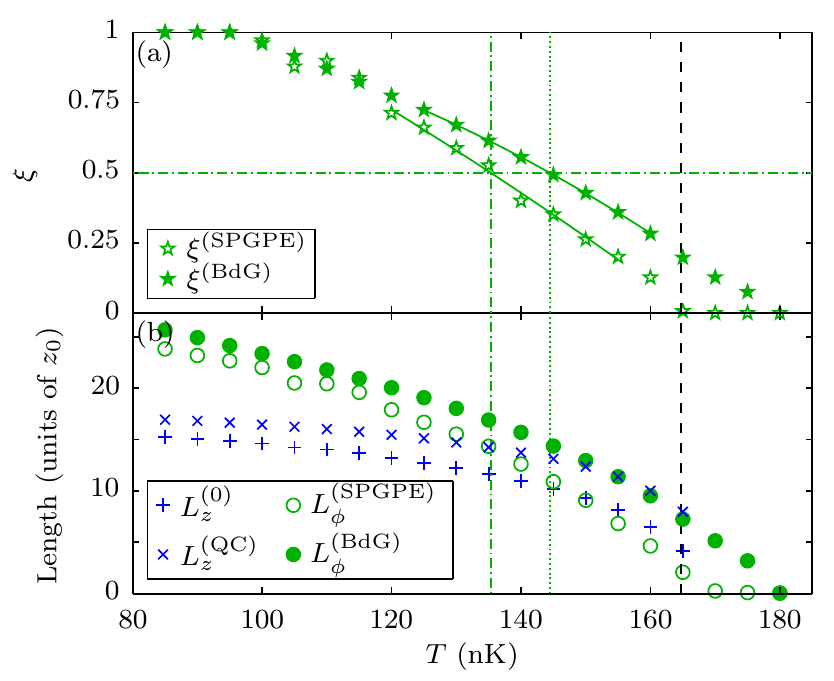}
	\end{center}
	\caption{\small (Color online) (a) Fitting parameter $\xi$, which characterizes the functional form of $g^{(1)}(-z,z)$ (see text), in SPGPE (open stars) and BdG \cite{Petrov2001a} (filled stars) calculations.  The horizontal dot-dashed line indicates the condition $\xi=0.5$, by which we define the crossover temperature.  The solid lines are smooth lines of best fit used to identify the intersection with $\xi=0.5$.  (b) Phase coherence length $L_{\phi}$ obtained from SPGPE (open circles) and BdG \cite{Petrov2001a} (filled circles) calculations, and fitted Thomas-Fermi lengths $L_z^{(0)}$ (pluses) and $L_z^{(\mathrm{QC})}$ (crosses) of the condensate and quasicondensate, respectively.  Vertical lines through both panels indicate the crossover temperature $T_\phi$ obtained from SPGPE (vertical dot-dashed line) and BdG \cite{Petrov2001a} (vertical dotted line) calculations, and the critical point identified using $C_B$ (vertical dashed line).  
	}  
	\label{fig:LphiGness}
\end{figure}

In Fig.~\ref{fig:LphiGness}(a) we plot the temperature dependence of $\xi$ (open stars), and note that the characteristic value $\xi=0.5$ of the crossover (horizontal dot-dashed line) is reached at a temperature $T_\phi = 135\;\nK$ (vertical dot-dashed line).  We observe a gradual change in $\xi$ through the crossover, from $\xi=0$ above $\Tc$ (vertical dashed line) to $\xi=1$ far below $\Tc$, consistent with the gradual change in the shape of $g^{(1)}(-z,z)$ shown in Fig.~\ref{fig:g1density}.

In Fig.~\ref{fig:LphiGness}(b) we plot the temperature dependence of $L_\phi$ (open circles), along with the estimated Thomas-Fermi lengths of the condensate $L_z^{(0)}$ (pluses) and quasicondensate $L_z^{(\mathrm{QC})}$ (crosses).  The two Thomas-Fermi lengths are obtained by performing fits of the condensate and quasicondensate column densities, $\overline{n_0}(y,z) \equiv \int\!dx\,n_0(\rbf)$ and $\overline{n\QC}(y,z) \equiv \int\!dx\,n\QC(\rbf)$, respectively, to the Thomas-Fermi column density  
\begin{equation}\label{eq:TF_fit}
	\overline{n_{\mathrm{TF}}}(y,z) = \overline{n_{\mathrm{TF}}}(0,0)\mathrm{max}\Big\{ 0 , \Big(1- \frac{y^2}{L_{\perp}^2} - \frac{z^2}{L_z^2}\Big)^{\frac{3}{2}}\Big\},
\end{equation}
obtained by integrating the Thomas-Fermi density $n_{\mathrm{TF}}(\rbf)=\mathrm{max}\{0,(\mu-V_{\mathrm{ext}}(\rbf))/U_0\}$~\cite{Pethick2008a} over $x$.  Here we regard the peak column density $\overline{n_{\mathrm{TF}}}(0,0)$, transverse Thomas-Fermi length $L_{\perp}$, and axial Thomas-Fermi length $L_z$ as independent fitting parameters~\cite{NoteH}.  Previous studies of this system and the related one-dimensional case loosely distinguish the quasicondensate and true condensate regimes by the size of the coherence length as compared to the Thomas-Fermi length of the quasicondensate~\cite{Petrov2000a,Petrov2001a,Proukakis2006b}.  Indeed we find here that at  temperatures in the range $T_\phi < T < \Tc$, the coherence length $L_\phi$ is smaller than the axial length of the quasicondensate $L_z^{(\mathrm{QC})}$, whereas the coherence length $L_\phi$ increasingly exceeds $L_z^{(\mathrm{QC})}$ as the temperature decreases below $T_\phi$.  We note in particular that equality of the two lengths ($L_\phi=L_z^{(\mathrm{QC})}$) appears to coincide almost exactly with the crossover temperature as defined by the condition $\xi=0.5$.

We now compare the crossover temperature obtained from our SPGPE simulations with the predictions of the BdG theory of Petrov \emph{et al.}~\cite{Petrov2001a}.  The authors of Ref.~\cite{Petrov2001a} derive an approximate expression
\begin{align}
	\langle [\delta\hat{\phi}(z,z')]^2 \rangle = & \; \frac{4\kB T \mu(N_0)}{15 N_0 (\hbar\omega_z)^2} \sum_{j=1}^\infty \frac{(j+2)(2j+3)}{j(j+1)(j+3)} \label{eq:dphi2Petrov} \\
	& \!\! \times \left[P_j^{(1,1)}\!\!\left(\frac{z}{L_z(N_0)}\right) - P_j^{(1,1)}\!\!\left(\frac{z'}{L_z(N_0)}\right) \right]^2 \!\! , \nonumber
\end{align}
for the variance of the phase difference between axial points $z$ and $z'$, where $\mu(N_0)$ and $L_z(N_0)$ are the Thomas-Fermi chemical potential and associated axial Thomas-Fermi length~\cite{Pethick2008a}, respectively, of a pure condensate of population $N_0$ [we use the ideal-gas estimate of the condensate population $N_0(T)$ --- see Sec.~\ref{sec:params}], and the $P_j^{(1,1)}$ are Jacobi polynomials.  Using Eq.~\reff{eq:dphi2Petrov} we calculate the symmetrically evaluated coherence function
\begin{equation}
	g^{(1)}(-z,z) = e^{-\frac{1}{2}\langle [\delta\hat{\phi}(-z,z)]^2\rangle},
\end{equation}
for each temperature $T$ we simulated with the SPGPE.  We then proceed to determine values of $\xi$ and $L_\phi$ by fitting $g^{(1)}(-z,z)$ with Eq.~\reff{eq:phase_fit}, and plot these quantities in Fig.~\ref{fig:LphiGness} (filled stars and filled circles, respectively) where they may be compared to our SPGPE results.

We find that the BdG model predicts the crossover criterion $\xi=0.5$ to occur at a temperature $T_\phi = 144\;\nK$ (vertical dotted line), which we note lies somewhat above the temperature at which it occurs in our SPGPE calculations (vertical dot-dashed line).  Proukakis~\cite{Proukakis2006a,Proukakis2006b} has shown that in true one-dimensional systems, the density fluctuations neglected in the treatment of Refs.~\cite{Petrov2000a,Petrov2001a} can erode phase coherence in the system, and thus push the crossover to lower temperatures.  We note that in addition, our 3D classical-field model includes fluctuations of the field along the transverse (tight) axes of the trap, which are neglected in the model of Ref.~\cite{Petrov2001a}, and may further reduce the phase coherence in the system.  Nevertheless, our quantitative results show that the analysis of Petrov~\emph{et al.}~\cite{Petrov2001a} captures the essential physics of the quasicondensate regime of degenerate Bose gases in elongated 3D traps.

\section{Conclusions} \label{sec:concl}
We have studied the correlations of a weakly interacting Bose gas in an elongated 3D harmonic trapping geometry, using a grand-canonical classical-field method.  Our investigations spanned temperatures ranging from the low-temperature true-condensate regime, through the crossover to the phase-fluctuating quasicondensate regime, to the normal phase of the gas at high temperatures.  We characterized the onset of condensation in the system using two independent measures:  the Binder cumulant quantifying number fluctuations in the Penrose--Onsager condensate orbital, and the suppression of the effective two-body interaction strength by many-body processes, as encoded by the anomalous thermal density of the field.  We found that both measures indicated that the transition to condensation occurs at a temperature slightly below the ideal-gas critical temperature for this geometry, as is the case in more standard 3D geometries \cite{Davis2006a,Smith2011a}.

However, our results show that the system remains in a strongly phase-fluctuating quasicondensate regime until significantly lower temperatures, as previously predicted for such an anisotropic geometry \cite{Petrov2001a}.  We explained that the quasicondensate phase should fundamentally be understood in terms of partial Bose condensation of the field, but that the large phase fluctuations in this regime imply that the noncondensed component of the gas cannot be understood in terms of Gaussian (Hartree-Fock-Bogoliubov) fluctuations.

We identified a temperature characteristic of the condensate--quasicondensate crossover, based on a qualitative change in the functional form of the non-local first-order coherence function, and found that this qualitative change occurs at roughly the same temperature at which the phase-coherence length equals the axial length of the quasicondensate.  The crossover temperature we find is somewhat lower than that predicted by an approximate Bogoliubov--de~Gennes model of phase fluctuations in the system~\cite{Petrov2001a}, consistent with the expectation that density fluctuations of the field have an additional deleterious effect on phase coherence~\cite{Proukakis2006a,Proukakis2006b}.

The pseudo-low-dimensional system we have considered is conceptually simpler than true low-dimensional systems, in that the appearance of the condensate, which here underlies the quasicondensate regime of the gas, is fundamentally due to a proper thermodynamic phase transition; i.e., it is expected to persist in the thermodynamic limit.  By contrast, in true low-dimensional systems, a significantly occupied orbital that can meaningfully be identified as a condensate may occur simply because of finite-size effects.  Nevertheless, the tools used in our characterization of the role of the condensate in the phase-fluctuating quasicondensate regime of this comparatively straightforward geometry offer to help elucidate the role of the Penrose-Onsager ``condensate'' in truly (quasi-)low-dimensional inhomogeneous Bose systems.

\begin{acknowledgments}
We acknowledge helpful discussions with J.~Carrasquilla, S.~P.~Cockburn, K.~V.~Kheruntsyan, and N.~P.~Proukakis, and thank U.~R.~Fischer for bringing Refs.~\cite{Fischer2002a,Fischer2005a} to our attention. This work was supported by the Australian Research Council through the Discovery Projects program (DP1094025, DP110101047).  M.C.G. acknowledges financial support from NSERC, Endeavour IPRS, and the University of Queensland.  T.M.W. acknowledges the hospitality of the KITP at UCSB, where this research was supported in part by NSF Grant No. PHY11-25915. 
\end{acknowledgments}

\appendix

\section{Simulation parameters}\label{apx:cutoff} 
\begin{figure*}[!t]
	\begin{center}
	\includegraphics{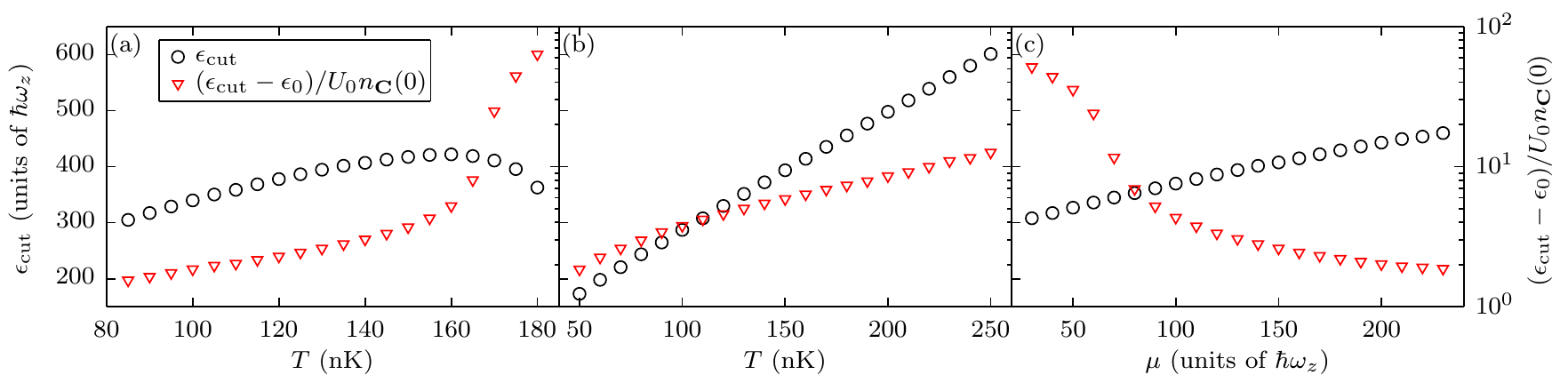}
	\end{center}
	\caption{\small (Color online) Energy cutoff $\epsilon_\mathrm{cut}$ (circles) and ratio $(\epsilon_\mathrm{cut} - \epsilon_0)/U_0n\C(0)$ of cutoff and interaction energies (triangles) for simulations with (a) varying $T$ at constant $N$, (b) varying $T$ at constant $\mu$, and (c) varying $\mu$ at constant $T$.}
	\label{fig:EcutNcut}
\end{figure*}

In choosing simulation parameters we must first ensure that our system satisfies the physical requirements to exhibit elongated 3D quasicondensate behavior over an appreciable temperature regime.  Using the estimates of Petrov \emph{et al.}~\cite{Petrov2001a} for $\Tc$ and $T_\phi$ in the limit $N_0\approx N$, we obtain the approximate scaling 
\begin{equation}
	\frac{T_\phi}{\Tc} \propto \left( \frac{\hbar^3 N^4 \omega_z^{19}}{m^3 a^6 \omega_\perp^{22}} \right)^{\frac{1}{15}}.
\label{eq:TphiTc}
\end{equation}
It can be seen that this ratio scales weakly with atom number, mass, and scattering length, and that its strongest dependence is on the frequencies $\omega_\perp$ and $\omega_z$ of the trapping potential.  The (relative) temperature range spanned by the quasicondensate regime therefore increases with increasing aspect ratio $\omega_\perp/\omega_z$.  However, our choice of aspect ratio is limited by the requirements that $\omega_\perp$ is small enough for the system to remain three-dimensional at the crossover (i.e., that $\kB T_\phi,\,\mu \gtrsim \hbar\omega_\perp$), and that the value of $\omega_z$ is experimentally reasonable. 

We must also take care to choose parameters such that the validity conditions of our classical-field methodology~\cite{Blakie2008a} can be satisfied.  The first such condition is that the cutoff energy $\epsilon_{\mathrm{cut}}$ is high enough that all eigenmodes of the single-particle Hamiltonian that are strongly coupled to one another by interactions (i.e., those which contribute to the quasicondensate density) are included in the $\mathbf{C}$~region.  We therefore require that the cutoff energy satisfies   
\begin{equation}\label{eq:high_cutoff}
(\epsilon_\mathrm{cut} - \epsilon_0)/U_0n\C(0) \gtrsim 1,
\end{equation}
where $\epsilon_0=\hbar(\omega_\perp + \omega_z/2)$ is the ground-state energy of the single-particle Hamiltonian [Eq.~\reff{eq:Hsp}] and $n\C(0)$ is the central (peak) density of the $\mathbf{C}$-region field (cf. Ref.~\cite{Bisset2009}).  We note that for a 3D system with $\kB T,\,\mu \gtrsim \hbar\omega_\perp$, this condition automatically implies that the classical field will span multiple modes in the transverse dimensions. 

The second c-field validity criterion is that all modes in the $\mathbf{C}$~region have mean occupations $\gtrsim 1$.  This condition is, in general, at variance with Eq.~\reff{eq:high_cutoff}, as raising the energy cutoff $\epsilon_{\mathrm{cut}}$ to accommodate the effects of interactions introduces progressively higher energy --- and therefore increasingly sparsely populated --- modes into the $\mathbf{C}$~region.  Simultaneously satisfying both of these classical-field conditions down to low temperatures $T\lesssim T_\phi$, while maintaining quasicondensate behavior over a reasonably large temperature range, is more readily achieved in systems with a relatively small dimensionless interaction energy $\tilde{U_0} = 4\pi a\sqrt{m\omega_z / \hbar}$, and for this reason we have chosen to simulate $^{23}$Na atoms. 

For a system of $N=2\times 10^5$ atoms of $^{23}$Na, in a trap with frequencies $(\omega_\perp,\omega_z) = 2\pi\times (250,5)$~Hz, we obtain in our simulations a temperature ratio of $T_\phi/\Tc\approx0.82$ (see Secs.~\ref{sec:critpt}~and~\ref{sec:crosspt}).  These parameters allow us to properly satisfy the classical-field validity conditions over the temperature range of interest:  For each set of parameters we simulated, we chose the value of $\epsilon_\mathrm{cut}$ (circles in Fig.~\ref{fig:EcutNcut}) such that $n_j \equiv \langle |\alpha_j|^2 \rangle \geq 2$ for each mode $Y_j(\rbf)$ of the classical field.  The corresponding values of the ratio $(\epsilon_\mathrm{cut} - \epsilon_0)/U_0n\C(0) \gtrsim 3/2$ (triangles in Fig.~\ref{fig:EcutNcut}) are generally sufficient to encompass the quasicondensate density within the $\mathbf{C}$~region.

\section{Semiclassical Hartree-Fock}\label{apx:SCHF}
\begin{figure}
	\begin{center}
	\includegraphics{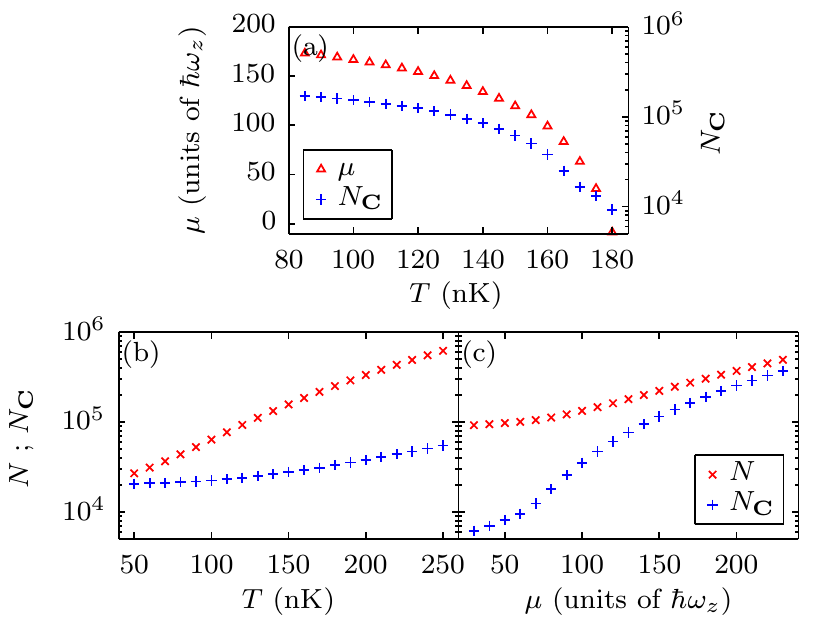}
	\end{center}
	\caption{\small (Color online) $\mathbf{C}$-region population $N\C$ (pluses), and semiclassical Hartree-Fock calculations of total ($\mathbf{C}$-region plus $\mathbf{I}$-region) atom number $N(\mu,T)$ (crosses) and chemical potential $\mu(N,T)$ (triangles). Quantities are shown for (a) varying $T$ at constant $N$, (b) varying $T$ at constant $\mu$, and (c) varying $\mu$ at constant $T$.}
	\label{fig:NmuTNc}
\end{figure}

We calculate equilibrium values of $N(\mu,T)$ and $\mu(N,T)$ (plotted in Fig.~\ref{fig:NmuTNc}) using semiclassical Hartree-Fock theory as outlined in Refs.~\cite{Goldman1981a,Huse1982a,Bagnato1987a,Oliva1989a,Shi1997a,Shi1997b,Giorgini1997b}. For a given choice of $\mu$ and $T$, the density of the thermal cloud $n_{\mathbf{I}}(\rbf)$ is calculated by integrating the Bose-Einstein distribution over momentum~\cite{Bradley2008a}, 
\begin{equation}\label{eq:nth}
   n_{\mathbf{I}}(\rbf)=\int_{|\pbf|\ge |\pbf_{\mathrm{min}}(\epsilon_{\mathrm{cut}},\rbf)|} \frac{d\pbf}{(2\pi\hbar)^3} \left\{e^{\left[\epsilon(\rbf,\pbf)-\mu\right]/\kB T}-1\right\}^{-1},
\end{equation}
where the semiclassical energy is 
\begin{equation}\label{eq:Esp}
    \epsilon(\rbf,\pbf)=\frac{p^2}{2m}+V_{\mathrm{ext}}(\rbf)+2U_0\left[n\C(\rbf)+n_{\mathbf{I}}(\rbf)\right],
\end{equation}
and $\pbf_{\mathrm{min}}$ is implicitly defined by $\epsilon(\rbf,\pbf_{\mathrm{min}}) = \epsilon_{\mathrm{cut}}$, ensuring that the integration over momentum is restricted to the $\mathbf{I}$ region. The semiclassical energy includes the mean-field potential from both the $\mathbf{C}$-region density $n\C(\rbf)$ and the above-cutoff density $n_{\mathbf{I}}(\rbf)$~\cite{Bradley2008a}. Equations~\reff{eq:nth}~and~\reff{eq:Esp} must therefore be solved self-consistently to determine the $\mathbf{I}$-region density profile for specified values of $\mu$ and $T$. Spatial integration of the total density then gives the total atom number, $N(\mu,T)=\int\!d\rbf\, [n\C(\rbf)+n_{\mathbf{I}}(\rbf)]$.  

To perform SPGPE simulations at fixed $N$ and varying $T$, we first make initial estimates for the chemical potential $\mu$ and the cutoff energy $E_\mathrm{cut}$.  In the case that $T>T_c^0$, we take the (negative) value of $\mu$ to be that for which $\int_0^\infty d\epsilon \,g(\epsilon)/[\exp\left(\beta(\epsilon-\mu)\right)-1] = N$, where $g(\epsilon)$ is the ideal-gas density of states appropriate to the elongated trap.  In the case that $T<T_c^0$, we use the Thomas-Fermi expression for $\mu(N_0)$, where the condensate population $N_0$ is given by the ideal-gas expression $N_0=N[1-(T/T_c^0)^3]$.  
We then take the cutoff energy $E_\mathrm{cut}$ to be the solution of $[\exp\left(\beta(E_\mathrm{cut}-\mu)\right)-1]^{-1}=N_\mathrm{cut}$, where we require $N_\mathrm{cut}=2$.  
An improved estimate for $\mu$ is found by inverting an approximate form for $N(\mu,T)$ obtained by assuming a Thomas-Fermi condensate density for the c-field density $n\C(\rbf)$ and calculating $n\I(\rbf)$ from Eqs.~\reff{eq:nth}~and~\reff{eq:Esp} with our estimate for $E_\mathrm{cut}$.  Our final value for $E_\mathrm{cut}$ is obtained by adding the average mean-field energy shift experienced by the highest energy single-particle eigenmodes $Y_n(\rbf)$ in the $\mathbf{C}$ region to our initial estimate for the cutoff energy.  We then evolve the SPGPE to find the equilibrium c-field density $n\C(\rbf)$ corresponding to the improved estimate $\mu$, and self-consistently solve Eqs.~\reff{eq:nth}~and~\reff{eq:Esp} to find the above-cutoff density $n\I(\rbf)$, and thus the total atom number $N$.  By iteratively adjusting $\mu$ and recalculating $n\C(\rbf)$ and $n\I(\rbf)$, $N$ can then be converged toward the desired value.  In practice, however, we find that only a single iteration is required to obtain convergence to within $1\%$ of the target value of~$N$.

\section{Fixed chemical potential and fixed temperature results}\label{apx:fixed_mu_and_fixed_T}
\begin{figure*}[!p]
	\begin{center}
	\includegraphics{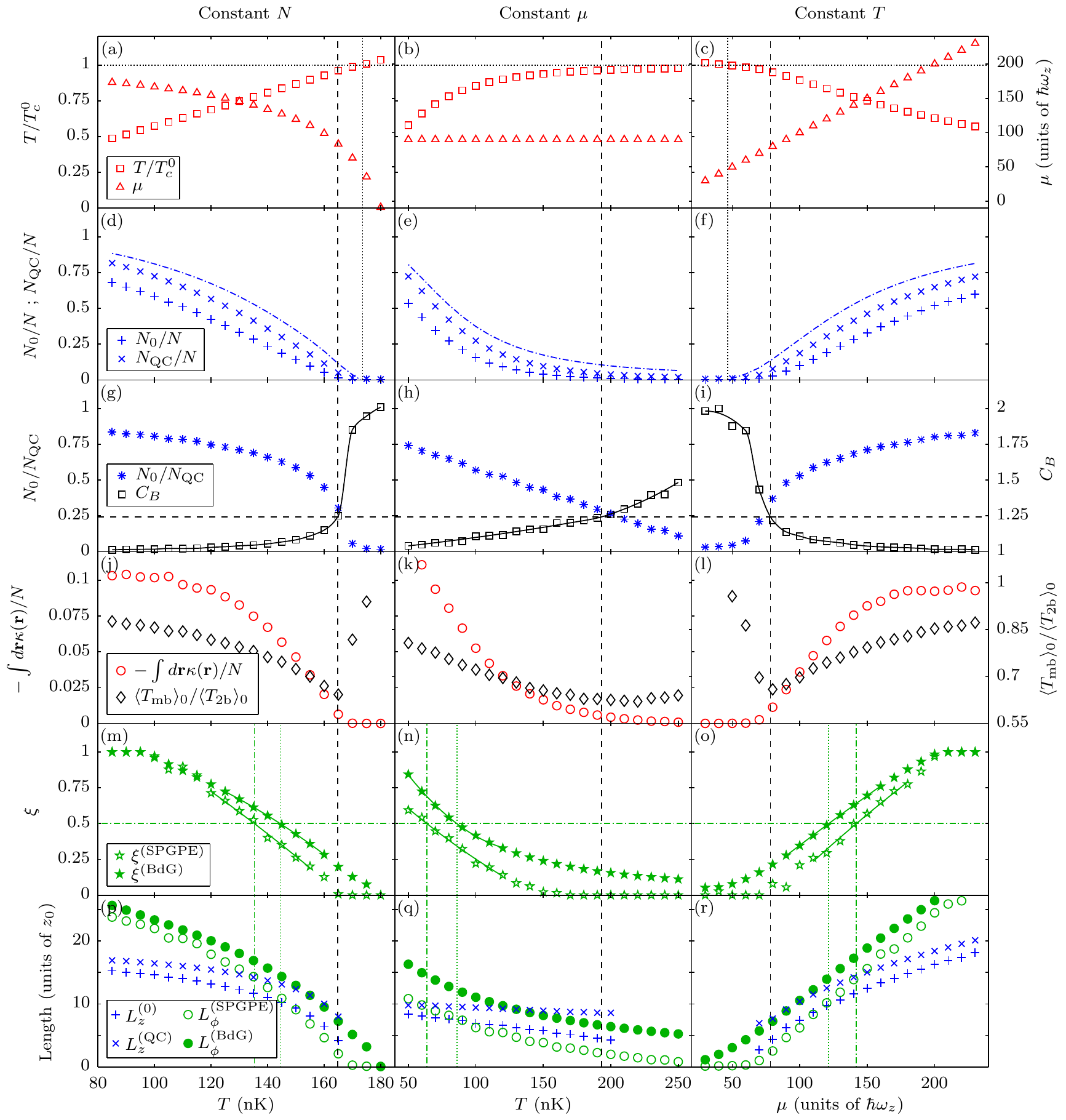}
	\end{center}
	\caption{\small (Color online) Properties of the system evaluated at varying $T$ and fixed $N$ (a,d,g,j,m,p), varying $T$ and fixed $\mu$ (b,e,h,k,n,q), and varying $\mu$ and fixed $T$ (c,f,i,l,o,r):  (a-c) Temperature $T$ (squares) and chemical potential $\mu$ (triangles). (d-f) Condensate fraction $N_0/N$ (pluses) and quasicondensate fraction $N\QC/N$ (crosses).  For reference the condensate fraction of the ideal Bose gas, and the ideal-gas critical temperature (chemical potential) are also indicated (dot-dashed line and vertical dotted line, respectively)~\cite{NoteI}.  (g-i) Ratio $N_0/N\QC$ of condensate and quasicondensate populations (asterisks) and generalized Binder cumulant $C_B$ (squares).  The horizontal dashed line represents the critical value $(C_B)_\mathrm{crit}$ for the Binder cumulant, and the vertical dashed line indicates the inferred estimate of the critical temperature.  The solid line is a smooth line of best fit used to identify the intersection with $(C_B)_\mathrm{crit}$.  (j-l) (Negative of the) integrated anomalous thermal density (circles) and effective suppression of scattering between condensate atoms due to associated processes (diamonds).  (m-o) Fitting parameter $\xi$, which characterizes the functional form of $g^{(1)}(-z,z)$ (see text), for SPGPE (open stars) and BdG \cite{Petrov2001a} (filled stars) calculations.  The horizontal dot-dashed line indicates the condition $\xi=0.5$, by which the crossover is characterized for SPGPE (vertical dot-dashed line) and BdG \cite{Petrov2001a} (vertical dotted line) calculations.  The solid lines are smooth lines of best fit used to identify the intersection with $\xi=0.5$.  (p-r) Phase coherence length $L_{\phi}$ for SPGPE (open circles) and BdG \cite{Petrov2001a} (filled circles) calculations, alongside fitted Thomas-Fermi lengths $L_z^{(0)}$ (pluses) and $L_z^{(\mathrm{QC})}$ (crosses) of the condensate and quasicondensate, respectively.  
	} 
	\label{fig:TcCBN0NQC}
\end{figure*}

In Fig.~\ref{fig:TcCBN0NQC} we present, in addition to the behavior of the system at constant particle number $N=2\times10^5$ over the temperature range $T = 85$ -- $180\;\nK$ discussed in Sec.~\ref{sec:results}, two additional sets of results for the field correlations:  those obtained over a range of temperatures  $T = 50$ -- $250\;\nK$ at a constant chemical potential $\mu = 90\hbar\omega_z$, and those obtained over a range of chemical potentials $\mu = 30$ -- $230\hbar\omega_z$ at a constant temperature $T = 135\;\nK$.

Figures~\ref{fig:TcCBN0NQC}(a)--\ref{fig:TcCBN0NQC}(c) show the system temperature (squares) as a fraction of the ideal gas critical temperature Eq.~\reff{eq:Tc0} [which itself varies with the varying total atom number in the data sets corresponding to Figs.~\ref{fig:TcCBN0NQC}(b) and~\ref{fig:TcCBN0NQC}(c)], and the system chemical potential (triangles).  Figures~\ref{fig:TcCBN0NQC}(d)--\ref{fig:TcCBN0NQC}(f) show the condensate fraction (pluses) and quasicondensate fraction (crosses), together with the ideal-gas result (dot-dashed line) for the condensate fraction at the corresponding values of $N$ and $T$.  In all three cases, the condensate and quasicondensate exhibit similar trends, increasing smoothly with decreasing $T$ (increasing $\mu$) from near zero at the highest temperatures (smallest chemical potentials) considered.  Moreover, the quasicondensate is in general significantly larger than the condensate, as shown explicitly by the ratio $N_0/N\QC$ (asterisks) in Figs.~\ref{fig:TcCBN0NQC}(g)--\ref{fig:TcCBN0NQC}(i).  The data also consistently show that $N_0$ and $N\QC$ are both suppressed below the corresponding ideal-gas predictions for the condensate occupation.  

In Figs.~\ref{fig:TcCBN0NQC}(g)--\ref{fig:TcCBN0NQC}(i) we plot, as in Fig.~\ref{fig:Nconst}(b), the Binder cumulant $C_B$ (squares) and observe that the estimate of the critical point obtained from the criterion $C_B=(C_B)_{\mathrm{crit}}$ uniformly indicates a lowering of the critical temperature, or correspondingly, a raising of the critical chemical potential (or density)~\cite{NoteI}.  This estimate of the critical point (vertical dashed line) is seen in all three cases to be reasonably consistent with the maximal suppression of the effective many-body $T$~matrix indicated in Figs.~\ref{fig:TcCBN0NQC}(j)--\ref{fig:TcCBN0NQC}(l) (diamonds).  Furthermore, we note that the condensate-quasicondensate population ratio $N_0/N\QC$ [asterisks in Figs.~\ref{fig:TcCBN0NQC}(g)--\ref{fig:TcCBN0NQC}(i)] assumes a value $N_0/N\QC\approx0.3$ at the identified critical point in all three data sets. 

In Figs.~\ref{fig:TcCBN0NQC}(m)--\ref{fig:TcCBN0NQC}(o) we plot the fitting parameter $\xi$ (open stars), which characterizes the functional form of $g^{(1)}(-z,z)$ [see Eq.~\reff{eq:phase_fit}], and identify the crossover temperature or chemical potential (vertical dot-dashed line) according to the condition $\xi=0.5$ (horizontal dot-dashed line).  In all three data sets, our results indicate a crossover to true condensation at a temperature (chemical potential) significantly lower (higher) than both the estimated critical point (vertical dashed line) and the crossover temperature obtained from the BdG model of Petrov~\emph{et al.}~\cite{Petrov2001a} (the corresponding $\xi$ values and crossover temperature are indicated by filled stars and a vertical dotted line, respectively).  Moreover, we see in Figs.~\ref{fig:TcCBN0NQC}(p)--\ref{fig:TcCBN0NQC}(r) that the intersection of the phase-coherence length $L_\phi$ (open circles) and Thomas-Fermi length $L_z^{(\mathrm{QC})}$ of the quasicondensate~\cite{NoteH} (crosses) coincides almost exactly with the crossover temperature (vertical dot-dashed line) in each of the three cases considered.

\bibliographystyle{prsty}

\begin{thebibliography}{100}
\bibitem{NoteH}
{We do not present the results of fits to the condensate and quasicondensate
  densities for temperatures $T \gtrsim T_c$ (or chemical potentials $\mu
  \lesssim \mu_c$), where the two densities are small and deviate strongly from
  a Thomas-Fermi profile.}
\bibitem{NoteI}
{We use the total atom number (Appendix~\ref{apx:SCHF}) to calculate the
  ideal-gas critical temperature [Eq.~(\ref{eq:Tc0})], chemical potential, and
  condensate fraction. We note that the ideal-gas transition is of course never
  reached at fixed $\mu>\epsilon_0=\hbar(\omega_\perp+\omega_z/2)$ (the
  zero-point energy of the trap), and the ratio $T/T_c^0$
  [Fig.~\ref{fig:TcCBN0NQC}(b)] approaches unity asymptotically as $T\to\infty$
  in this case.}
\end{thebibliography}

\end{document}